# Elastic deformation of optical coherence tomography images of diabetic macular edema for deep-learning models training: how far to go?

Daniel Bar-David[1, *]; Laura Bar-David[2, *]; Yinon Shapira[3,4]; Rina Leibu[2]; Dalia Dori[2]; Ronit Schneor[1]; Anath Fischer[1]; Shiri Soudry[2,5,6]

[1] Faculty of Mechanical Engineering, Technion Israel Institute of Technology, Haifa, Israel

[2] Department of Ophthalmology, Rambam Health Care Campus, Haifa, Israel

[3] Discipline of Ophthalmology and Visual Science, University of Adelaide, Adelaide, South Australia, Australia

[4] Department of Ophthalmology, Royal Adelaide Hospital and South Australian Institute of Ophthalmology, Adelaide, South Australia, Australia

[5] Clinical Research Institute at Rambam, Rambam Health Care Campus, Haifa

[6] Ruth and Bruce Faculty of Medicine, Technion Israel Institute of Technology, Haifa

[*]These authors contributed equally as first authors.

**Financial support:** This study received funding from the Israeli Ministry of Health Koppel Grant number 2028211. The sponsor or funding organization had no role in the design or conduct of this research.

**Conflict of interest:** No conflicting relationship exists for any author

**Running head**: Elastic deformation of OCT for data augmentation

**Keywords:** OCT, DME, Deep Learning, data augmentation, elastic deformation

**Abbreviations/acronyms:**
CAD (Computer-Aided Diagnosis),
CNN (Convolutional neural networks),
CRT (Central Retinal Thickness),
DL (Deep Learning),
DME (Diabetic Macular Edema),
NPDR (Non Proliferative Diabetic Retinopathy),
PDR (Proliferative Diabetic Retinopathy),
OCT (Optical Coherence Tomography)
**Word count:** 3350

**Abstract**

**Purpose**: To explore the clinical validity of elastic deformation of optical coherence tomography (OCT) images for data augmentation in the development of deep-learning model for detection of diabetic macular edema (DME).

**Methods**: Prospective evaluation of OCT images of DME (n=320) subject to elastic transformation, with the deformation intensity represented by ($\sigma$). Three sets of images, each comprising 100 pairs of scans (100 original & 100 modified), were grouped according to the range of ($\sigma$), including low-, medium- and high-degree of augmentation; ($\sigma$ =1-6), ($\sigma$ =7-12), and ($\sigma$ =13-18), respectively. Images subject to extremely high augmentation ($\sigma$ =19-24) served as control. Three retina specialists evaluated all datasets in a blinded manner and designated each image as 'original' versus 'modified'. The rate of assignment of 'original' value to modified images (false-negative) was determined for each grader in each dataset.

**Results**: The false-negative rates ranged between (71-77%) for the low-, (63-76%) for the medium-, and (50-75%) for the high-augmentation categories. The corresponding rates of correct identification of original images ranged between (75-85%, p>0.05) in the low-, (73-85%, p>0.05 for graders 1 & 2, p=0.01 for grader 3) in the medium-, and (81-91%, p<0.005) in the high-augmentation categories. In the control set, the rates of labelling modified images as 'original' were 20-65%. In the subcategory ($\sigma$ =7-9) the false-negative rates were (83-93%), whereas the rates of correctly identifying original images ranged between (89-99%, p>0.05 for all graders).

**Conclusions**: Deformation of low-medium intensity ($\sigma$=1-9) may be applied without compromising OCT image representativeness in DME.

In the developed world, diabetic retinopathy is a leading cause of preventable blindness among the working age population [1]. Of an estimated 425 million people with diabetes worldwide, nearly 10% are afflicted with a vision-threatening disease, with diabetic macular edema (DME) being the leading etiology [2]. Left untreated, DME is associated with an increased risk for irreversible central visual loss [3], hence the importance of early detection and treatment [4]. This condition is characterized by abnormal retinal thickening, a cystic pattern of intraretinal fluid accumulation, and intraretinal lipid or protein deposition which can be detected on ophthalmoscopic examination of the macula. Optical coherence tomography (OCT)-derived measures are the standard of care in the diagnosis of DME and in monitoring of therapeutic effects [5]. The multitude of morphological information provided by OCT in eyes with DME has advanced our understanding of this medical condition, because it enabled better detection and quantification of the macular thickening [6]. More importantly, it demonstrated that macular edema extends to additional features beyond mere retinal fluid [7]. Indeed, DME is a complex clinical entity with various morphological characteristics that should be considered to choose the appropriate therapeutic approach and understand its potential benefits. The analysis of OCT and identification of its related pathological features is complex and requires highly trained retina experts. Manual interpretations are extremely time consuming, with variable repeatability and interobserver agreement.

Computer-Aided Diagnosis (CAD) systems can facilitate interpretation of medical images, with rising global interest. In recent years, deep-learning models have been applied in CAD, leading to meaningfully improved results and higher ability to automatically detect abnormalities on medical images [8]. Deep-learning, a class of machine-learning inspired by the neuronal layers that constitute the human brain, has generated a revitalization in the fields of artificial intelligence and computer-aided vision. It utilizes multiple layers of neural networks to receive, process and extract features with various levels of abstraction from the data input without the need of manual feature engineering. Convolutional neural networks (CNN) is a special type of neural networks which is typically applied for image analysis tasks because it preserves the spatial relationship between pixels. In the field of retinal imaging, several studies based on CNN have already shown good performance on OCT images to classify retinal diseases such as diabetic retinopathy [9,10] and age-related macular degeneration [11,12], and to identify their features [13,14]. It is expected that deep-learning methods will continue to be applied for the development of computational OCT-based diagnostic tools for DME and additional macular conditions.

One of the major challenges to the development of any deep-learning based image analysis is the need of large training datasets, which are sets of example images used to construct and fit the algorithm. In the case of supervised learning, annotated images are needed for the algorithm training. To date, only few large public OCT datasets from multiple imaging devices are widely available [15]. Moreover, the extraction of sufficiently-large dataset of macular images and human reading for their manual annotation is time-consuming and requires high-level expertise in retinal medicine and detailed interpretation of OCT imaging. Strategies have been developed to overcome this challenge and counter the effect of limited datasets comprising small numbers of annotated images, to enable effective training of CNN. Among these, data augmentation is commonly employed. Data augmentation provides an effective approach to artificially expand and diversify an existing dataset without the need to acquire new images, by applying transformations on original ones. Some of the most popular data augmentation approaches include basic transformations such as random flipping, rotating, scaling, shifting, noising, and others. These elemental conversions are vastly used given their proved efficacy in improving performance [16].

A higher-level technique for data augmentation involves the introduction of random elastic deformations, in which the shape, geometry, and size of the object can be modified, often in a complex manner (**figure 1**). Implementation of elastic deformations consists of several separate steps and therefore can be more variable than basic transformations. In medical imaging, living human objects are inherently subject to naturally-occurring transformations which can be extrapolated to elastic deformations for the purpose of data augmentation for training datasets. To date, however, the most appropriate application of elastic deformation on retinal images for training data has not been determined.

Here, we explored this approach in the development of deep-learning model for automated detection of DME in OCT images. Specifically, we set to determine the degree of elastic deformation which could be applied to OCT images from eyes with DME without compromising their realism or clinical aptness.

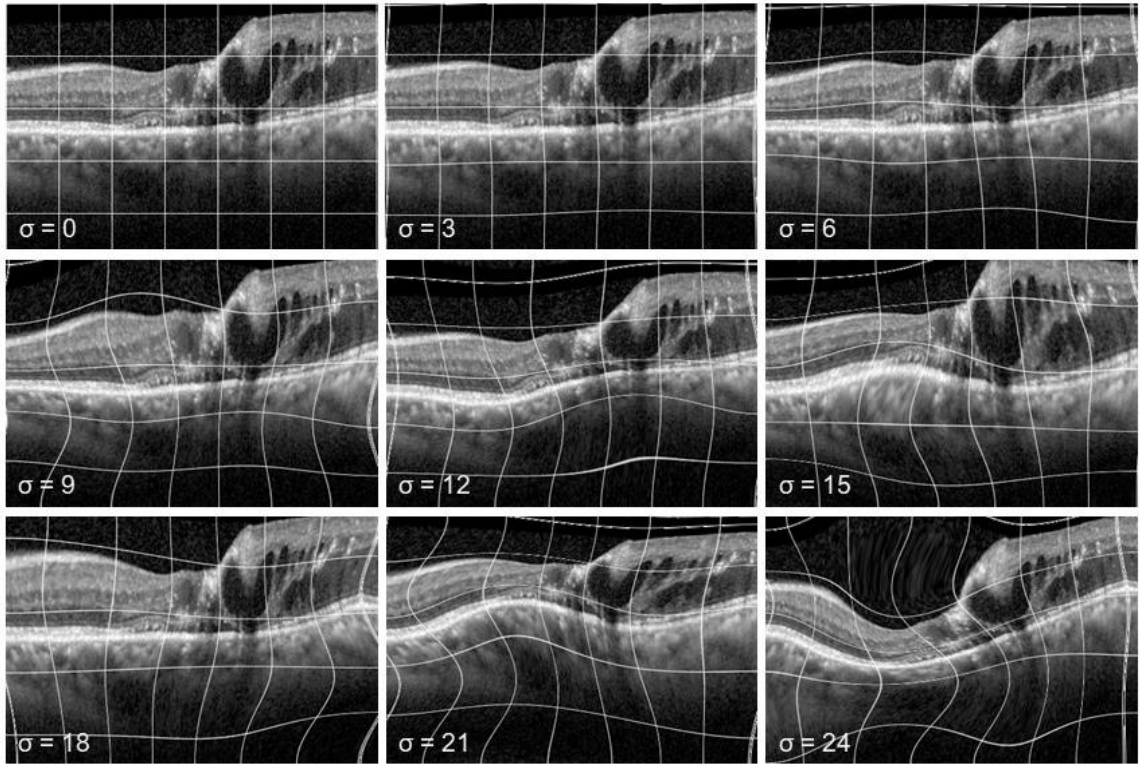

**Figure 1.** Elastic transformation of optical coherence tomography (OCT). An OCT image from an eye with diabetic macular edema (DME) was subjected to varying intensities of elastic deformation, ranging between σ=0 (no deformation) to σ=24. The grid is added to better visualize the configuration and magnitude of the applied transformation.

**Methods**

Dataset

The study protocol was approved by the Institutional Review Board of the Rambam Health Care Campus, Haifa, Israel, and adhered to the tenets of the Declaration of Helsinki. A database was gathered comprising macular spectral domain (SD)-OCT images from patients treated at the Retina Service of the Department of Ophthalmology, Rambam Health Care Campus, Haifa, Israel, between 2016 and 2019. Eligible participants for the study were patients affected by DME (**Table 1**) from whom the images were acquired as part of their routine clinical care. Images from 320 patients (320 eyes) were randomly selected for the study. All images were acquired with a HRA+OCT Spectralis OCT device (Heidelberg Engineering GmbH 69121 Heidelberg, Germany) using a 49-line raster macula scan. From each subject, a single cross-sectional macular (2D) image encompassing the foveal center was selected. All images had a resolution of 496 x 352 pixels and were collected in a de-identified manner.

| **Characteristic** | **Rate among subjects (n=320)** |
|---|---|
| **Age, mean (SD), years** | 73±5.7 |
| **Male** (%) | 47.6% |
| **Systemic comorbidities** | |
| Hypertension, hypercholesterolemia, ischemic heart disease (%) | 78.4% |
| Type II diabetes | 88.7% |
| HbA1C, % | 10.8±2.6 |
| **Ophthalmic data** | |
| OD (%) | 53.4% |
| OS (%) | 46.6% |
| Retinopathy grade | NPDR- 43.2%, PDR- 28.3% |
| CRT in study eye, mean (SD), μm | 361±102 |
| Center-involving DME (%) | 73% |

**Table 1**. Baseline characteristics of the study participants
NPDR: Non proliferative diabetic retinopathy, PDR: Proliferative diabetic retinopathy, CRT: Central retinal thickness.

Elastic deformation

Previously reported distortion approaches[17,18] seem improper for retinal imaging since they involve local displacement which can introduce overlap and discontinuity of the retinal layers. We therefore applied a more global elastic deformation which affects a relatively large area of the image, leaving the image overall smooth and continuous.

The elastic deformation process of a 2D OCT scan consists of three main steps.

First, a uniform 2D grid composed of n by m discrete cells is generated. Each cell is assigned with an arbitrary value which expresses the magnitude of deformation applied onto this region. The size of the grid was set empirically (m=n=3) in order to create a relatively global transformation such that the deformation of each cell affects a significant part of the image. The value of each cell $x_{mn}$ is randomly sampled from a normal distribution of mean μ=0 and a standard deviation σ experimentally set:

$$\chi \sim N(\mu = 0, \sigma^2)$$

This 3x3 grid fully covers the area of the 2D OCT scan.

The second step is to generate a displacement field (**figure 2**) that describes the elastic deformation between the original 2D OCT scan and the modified one. The displacement field consists of displacement vectors which define for all pixels of the 2D OCT scan the distance and direction from the initial position to the final position. The displacement vector is determined by using spline interpolation between the $x_{mn}$ values of the 3x3 grid, achieving a continuous and smooth displacement field $u(x, y)$.

Finally, the displacement field is applied on the 2D OCT scan where each pixel intensity is shifted to its new position by using bilinear interpolation.

$$P_m = P_o + u$$

where $u$ is the displacement field, and $P_m$ and $P_o$ are all the pixels positions in the modified and original image, respectively.

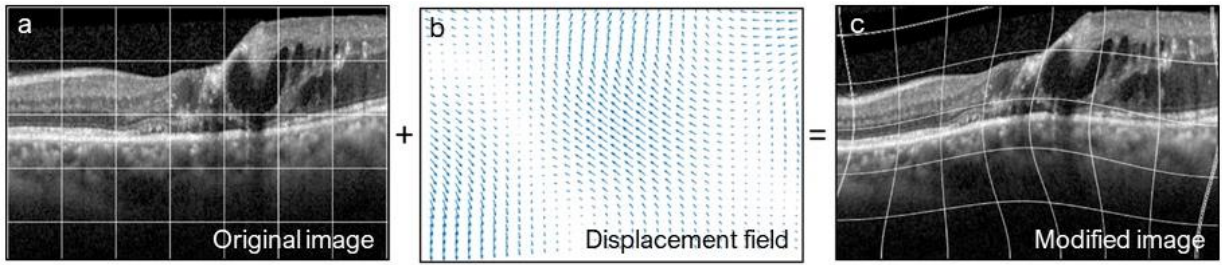

**Figure 2.** Applying realistic elastic deformation on OCT images. (a) an original OCT scan. (b) The displacement field consists of displacement vectors that define for each pixel in the original OCT image the distance and direction of displacement from the initial position to the final position. (c) The displacement vectors applied on the 2D OCT image result in the modified OCT image.

The σ value indicates the intensity of the deformation. As shown in **figure 1**, when the standard deviation σ is low, the randomly generated grid values are close to zero and consequently the displacement field is small resulting with low deformation. For increasing σ values, the randomly generated grid values are larger and consequently the displacement field is significant resulting with more readily apparent deformation.

Evaluation of the deformation

Three sets of images, each comprising 100 pairs of scans, i.e 100 original and 100 modified, were grouped according to the intensity of the elastic deformation applied. The range of the deformation (σ) selected for each category was 1-6, 7-12, and 13-18, for the low-, medium- and high-degree of augmentation, respectively (**figure 3**). A fourth set of images, including 20 pairs of images subject to extremely high augmentation σ range 19-24, served as control.

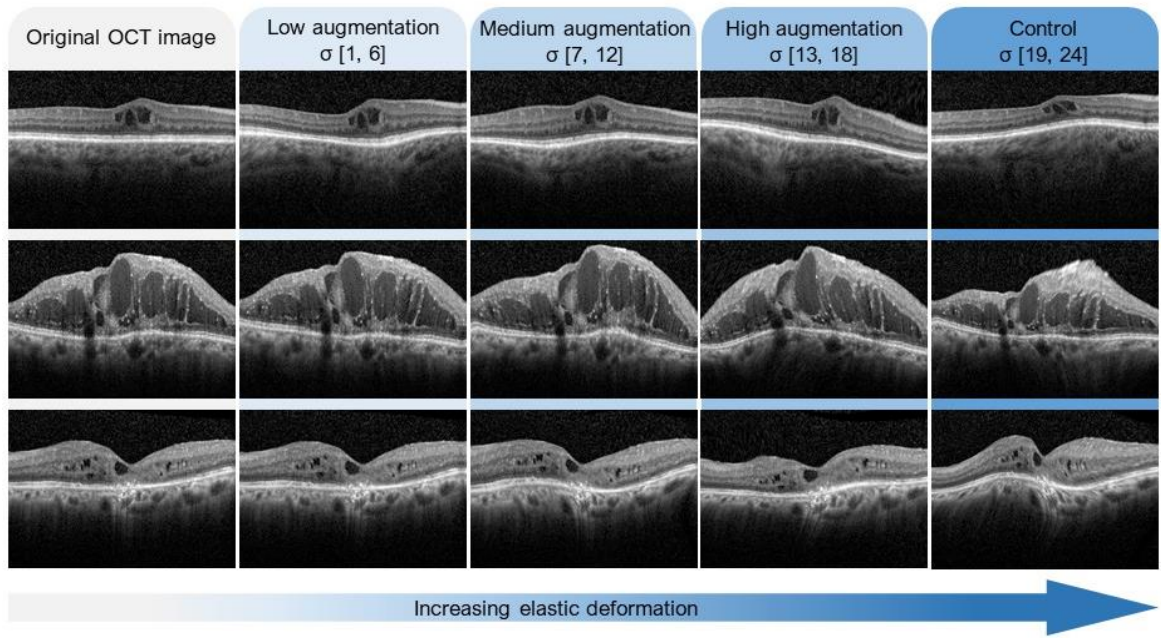

**Figure 3.** Categories of elastic transformation of optical coherence tomography (OCT) images of diabetic macular edema (DME). Four ranges of deformation intensity: low-, medium-, high-, and extreme-level of deformation, indicated by increasing σ values, were applied on original OCT images. The higher σ value is, the greater distortion in the image becomes more apparent.

Three retina specialists with an extensive clinical experience (more than 10 years) in the management of macular diseases and analysis of clinical OCT images were recruited as graders for an evaluation study.

In the first step of the study, each grader independently evaluated the 640 OCT images and was asked to determine whether they were original images or potentially deformed ones. To facilitate the process, a graphical user interface was implemented, where the image is displayed on the screen and the user can select one of two possible choices: *Original* or *Modified*. The 640 images from the four categories of deformation level were shuffled and displayed in a random order. The reviewers were able to go forward or backward without time limit to evaluate each image. To get the most objective answers, the reviewers were not provided with any information on the percentage of original scans in the dataset, nor were they informed on the degrees of elastic deformation employed.

Next, following examination of the results obtained at the first grading step, to refine the maximal deformation level which was deemed realistic by the clinical graders, a second study step was undertaken. A new range of data augmentation level (σ) was selected based on smaller bins of the intermediate range of deformations, namely (σ = 7-9) and (σ=10-11), and a different set of OCT images, including 100 pairs of images (100 original and 100 modified counterparts) was designed for each new subcategory. The 3 clinical readers independently graded the new datasets and were asked to determine whether each of the images could be acceptable as a representative image of DME, in their opinion.

Sample size calculation and Statistical analysis

Data were analyzed by the StatSoft Statistica software, version 10 (StatSoft, OK, USA). The following assumptions and criteria defined a "realistic" intensity of deformation:

1) A minimum of 60% rate of designation of augmented (modified) images as 'original' (false-negatives) served as a criterion to define a "realistic" intensity of deformation. This criterion corresponds to a delta of 10% from chance designation (50%) by the graders. That is, this rate corresponds to a higher chance of designating an augmented image as 'original' than labeling it as 'modified'.

2) The original (non-modified) images served as the 'standard' comparison group representing "real-life" non-modified images recognition. We assumed that at least 85% of the 'standard' images will be correctly assigned as 'original' (true-negatives), accounting for anticipated confusion error during the rating of the random images presented in the sets.

3) For each grader (and each set/category of deformation), a non-significant difference in true-negatives and false-negatives rates was mandatory in order to define a "realistic" intensity of deformation.

4) For each grader and each set, both criteria 1 and 3 had to be met. To this end a non-inferiority sample size calculation was performed yielding the following null hypothesis: If there is a true difference in favor of an 'original' image recognition [standard group] of 5%, then 192 images (96 images per group) are required to be 80% confident ($\beta=0.8$) that the upper limit of a one-sided 95% confidence interval ($\alpha=0.05$) will exclude a difference in favor of the standard group of more than 20%. Based on this assumption, 200 images were included in each set of deformation category (100 non-modified and 100 modified counterparts). Proportions of true-negatives and false-negatives were compared by chi-square test. A two-sided p value of less than 0.05 was considered significant.

**Results**

Deformation intensity evaluation

To evaluate the realistic value of the images displaying varying intensities of deformity, we compared the rate of labelling original images as 'original' and modified images as 'original' (false-negative) by the graders in each category of augmentation level. The readings obtained from the three experts in each category are summarized in **figure 4**.

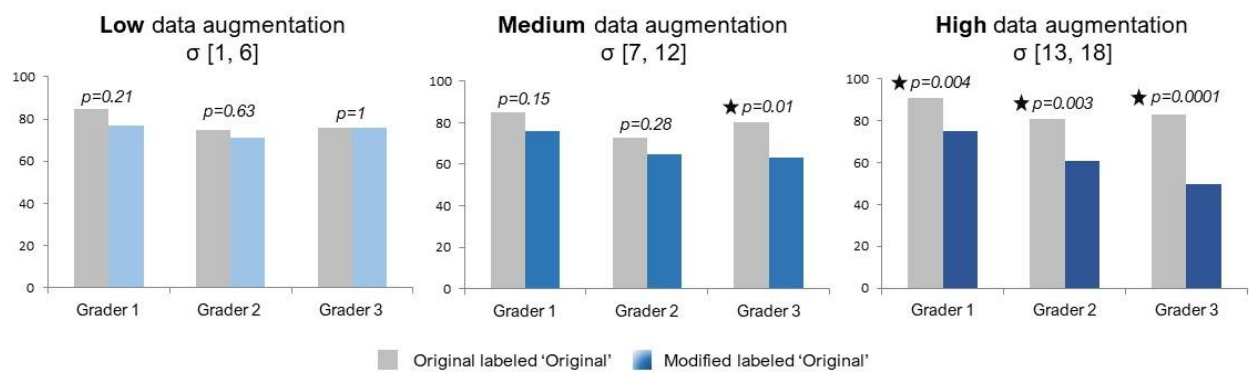

**Figure 4.** Pairwise comparison of the number of modified images labeled 'original' (blue columns) and the original images labeled 'original' (grey columns) by the three graders for the Low-, Medium-, and High- levels of image data augmentation.

Among the 3 graders, the rates of labelling modified images as 'original' (false-negative) (**Supplementary table 1**) ranged between (71-77%) for the category of low augmentation, (63-76%) for the medium augmentation, and (50-75%) for the high-augmentation category, indicating overall decreasing alleged realism of the modified images with higher level of distortion. In comparison, the corresponding rates of correctly identifying original images as 'original' (true-negative) ranged between (75-85%, $p>0.05$ for all graders) for low augmentation category, (73-85%, $p>0.05$ for graders 1 & 2, $p=0.01$ for grader 3) for the medium augmentation category, and (81-91%, $p<0.005$ for all graders) for high augmentation category. Thus, the frequency of proper identification of original images as 'original' remained similarly high for all categories of image distortion, whereas the realistic value of the modified images decreased with increasing deformation. Specifically, for the low augmentation category, there was no significant difference between the 'original' images true-negative and the 'modified images' false-negative rates for all graders **(figure 5)**, indicating that low levels of distortion did not compromise the apparent realism of the modified images.

For the medium category, only one grader (grader 3) was significantly less likely to designate modified images as 'original', whereas for the other 2 graders the frequency of labeling modified images as 'original' was comparable to the frequency of designating original images as 'original'. In contrast, for the high augmentation category, all graders showed a higher frequency of appropriately identifying original images as 'original' (true-negatives) than modified images as 'original' (false-negative).

In the control set (extremely high augmentation, σ range 19-24), the rates of labelling modified images as 'original' (false-negative) by the 3 graders were 20-65% compared to 75-100% (true-negative) rate for the 'original' counterparts (p≤0.008 for all graders).

Thus, application of low levels of elastic deformation (σ = 1-6) did not compromise the OCT image representativeness in the case of DME. Higher levels of deformation (σ ≥ 13) resulted in unrealistic images which were more readily interpreted as modified by the clinical graders.

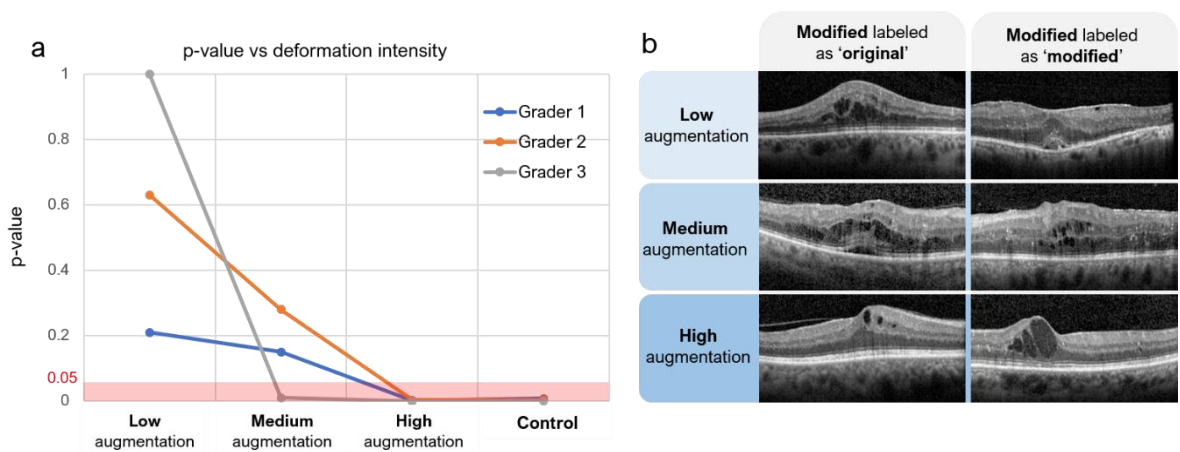

**Figure 5.** (a) Probability of detection of modified OCT images in the different categories of deformation intensity. For each grader, the calculated p-value for proper designation of original images as 'original' (true-negative) versus designation of modified images as 'original' (false-negative) is shown. When p-value=1, the grader labeled an equal number of original and modified images as 'original'. Smaller p-values indicate lower rates of designation of modified images as 'original'. P-values smaller than 0.05 (red line) denote a statistical significance that the grader identified a difference between original and modified images. (b) examples for images uniformly designated by all 3 graders in each category of augmentation.

Refining deformation intensity range

For the category of medium intensities of deformation (σ = 7-12), the frequency of detection of distortion varied among the graders, indicating inconclusive realism of the modified images. Therefore, we refined the maximal deformation level which could be interpreted as realistic within this range. Two new datasets consisting of smaller bins of the deformation values of the of medium range (σ = 7-9 and σ= 10-11) were selected. The experiment was reiterated using the same methodology outlined above (i.e., a set of 100 modified and 100 'original'/unmodified images for each new category). The results are presented in **Figure 6**.

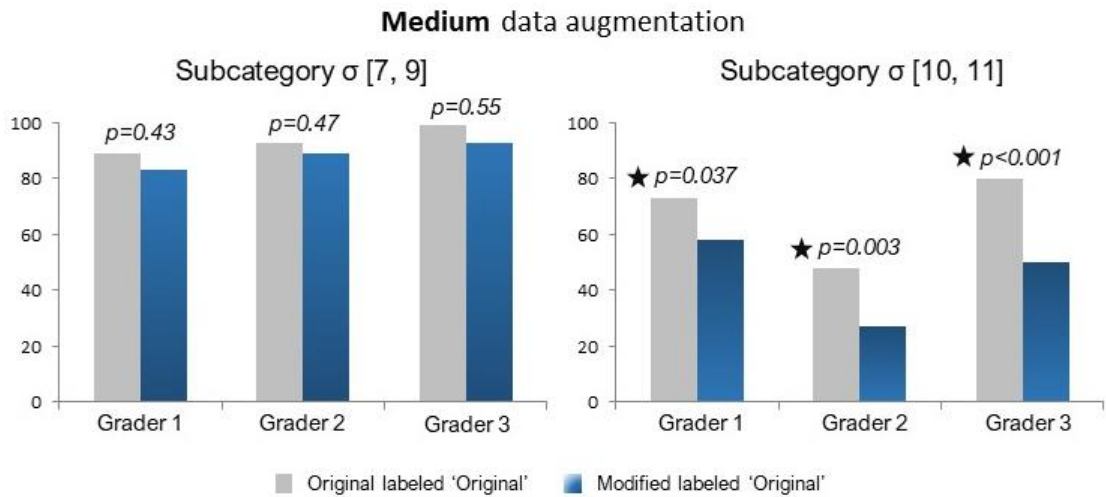

**Figure 6.** Pairwise comparison of the number of modified images labeled 'original' (blue columns) and the original images labeled 'original' (grey columns) by the three graders for the two new subcategories of the medium augmentation range (σ = 7-9 and σ =10-11).

For the lower subcategory (σ =7-9), the rates of labelling modified images as 'original' (**Supplementary table 2**) among the 3 graders ranged between (83-93%). In comparison, the corresponding rates of correctly identifying original images as 'original' ranged between (89-99%, p>0.05 for all graders). Hence, the frequency of correct identification of original images as 'original' (true-negative) was comparable to the rate of classification of modified images as 'original' (false-negative) for each grader (p=0.43-0.55 for all graders). In contrast, in the subcategory consisting of higher augmentation (σ =10-11), all 3 graders properly were significantly more likely to correctly identify modified images, indicating unrealistic clinical suitability.

Overall, the graders did not significantly detect modified images manifesting deformation in the range (σ = 0-9).

**Discussion**

The primary importance of data augmentation for the development of deep-learning models is to provide the algorithm with richer training datasets. In addition, by addressing the lack of large image datasets required for training, data augmentation can enhance our ability to meet one of the main methodological challenges currently restricting the development of deep-learning models in OCT data [19]. The deformation approach is a key method of image data augmentation which can facilitate the learning process and improve its performance. In this paper, we aimed to determine the maximal degree of elastic deformation which could be employed while maintaining the clinical realistic value of OCT images in the case of DME. To our knowledge, this is the first report exploring the quantitative parameters of elastic deformations of OCT macular images and validating the clinical representativeness of such transformations by retina specialists.

In practical application of data augmentation, it is important to utilize a method which potentially represents a clinical equivalent and maintains the realistic value of the medical images, because this approach can directly affect the algorithm output. The retina is a biological membrane with biomechanical properties allowing it to stretch and deform in response to various forces. Processes such as accumulation of intraretinal or subretinal fluid, traction exerted by the adjacent vitreous or epiretinal fibrocellular proliferation, can all lead to abnormalities in the shape and configuration of the retina. Moreover, the elastic qualities of the retina permit reversibility of such resulting deformations, allowing the affected tissue to regain its contour upon resolution of the adverse mechanism. DME often manifests with a spectrum of clinically observable changes in the geometry and structure of the macula which are pliable in nature. Thus, application of elastic transformation was inferred as a valid approach for data augmentation in the case of macular images from eyes with DME. To this end, it was interesting to note that although infrequently so, some of the original OCT images in our study were designated as 'modified' by all 3 graders. We postulate that naturally occurring DME-related deformations in the macular contour were interpreted in some cases as displaying "distortion" in the readers opinion and have accounted for such discrepancies.

We found that for low levels of deformation ($\sigma$ = 1-6), 3 clinical graders were unable to significantly distinguish between original and modified OCT images showing various morphological manifestations of DME. Therefore, this range of deformation can be applied without meaningfully compromising OCT image representativeness in the case of DME. Higher levels of deformation ($\sigma \geq 13$) resulted in unrealistic images which were frequently recognized as modified by the graders. For the range of medium intensities of deformation (σ = 7-12), the rate of proper detection of artificial distortion varied among the graders, indicating inconclusive realistic value of the modified images. To refine the range of deformation intensity according to clinical adequacy in the medium category, we adopted an approach comparable to the bisection method. The interval was bisected into two subintervals, i.e. (σ = 7-9) and (σ = 10-11). In the lower range of deformation intensities (σ = 7-9), modified images were interpreted as original at a similar rate to that of original images, suggesting that this range of deformation indeed did not interfere with the representativeness of OCT images. In contrast, in the higher subgroup of deformations (σ = 10-11) modified images were significantly detected by all 3 graders. We therefore concluded that the overall range of elastic deformation which could be applied to OCT images from eyes with DME without compromising their representativeness is (σ = 0-9).

Medical imaging entails a visual representation of a unique anatomy or function of organs and tissues. Each artificial transformation of the image can alter its realism and impede its clinical representativeness, thus potentially leading to nonrealistic or erroneous diagnosis. Such considerations become particularly relevant while employing elastic deformations as data augmentation techniques for training of deep-learning based algorithms. Our study involved three retina specialists with extensive experience in the interpretation of clinical OCT scans and their implementation in daily management of a wide-ranging population of patients with various macular disorders, including DME. Moreover, for the present study each grader evaluated over 640 images from a diverse group of patients manifesting a broad spectrum of disease characteristics and severity levels (**Table 1**). Thus, the comprehensive dataset employed in our study supports good generalizability of our results, suggesting that elastic deformation in the range (σ = 0-9) could be applied to OCT images from the general population of DME patients. Beyond this range of deformation, artificially modified images failed to represent morphological variability naturally evident such eyes.

Our results confirmed that application of elastic deformations on OCT images of DME is a valid strategy for data augmentation which did not affect their realistic value. We identified a range of intensity of elastic deformations which we suggest could be extrapolated for use in deep-learning based detection of additional retinal pathologies manifesting similar features and shared phenotypes. Future studies will further validate this assumption.

**Supplementary information**

| | LDA $\sigma \epsilon$ [1-6] (n=200) | | | MDA $\sigma \epsilon$ [7-12] (n=200) | | | HDA $\sigma \epsilon$ [13-18] (n=200) | | | Control $\sigma \epsilon$ [19-24] (n=40) | | |
|---|---|---|---|---|---|---|---|---|---|---|---|---|
| | Original (n=100) | Modified (n=100) | *p* value | Original (n=100) | Modified (n=100) | *p* value | Original (n=100) | Modified (n=100) | *p* value | Original (n=20) | Modified (n=200) | *p* value |
| Grader 1 | 85 | 77 | 0.21 | 85 | 76 | 0.15 | 91 | 75 | 4$^e$-4 | 20 | 13 | 8$^e$-3 |
| Grader 2 | 75 | 71 | 0.63 | 73 | 65 | 0.28 | 81 | 61 | 3$^e$-3 | 17 | 4 | 4$^e$-3 |
| Grader 3 | 76 | 76 | 1 | 80 | 63 | 0.01 | 83 | 50 | 1$^e$-4 | 15 | 4 | 1$^e$-3 |

**Supplementary Table 1.** Quantitative evaluation of the realistic value of the images by the three reviewers for the four categories of levels of the incorporated elastic deformation: Low-, Medium-, High- data augmentation and control.

| | Deformation intensity $\sigma \epsilon$ [7-9] (n=200) | | | Deformation intensity $\sigma \epsilon$ [10-11] (n=200) | | |
|---|---|---|---|---|---|---|
| | Original (n=100) | Modified (n=100) | *p* value | Original (n=100) | Modified (n=100) | *p* value |
| Grader 1 | 89 | 93 | 0.43 | 73 | 58 | 0.037 |
| Grader 2 | 93 | 89 | 0.47 | 48 | 27 | 0.003 |
| Grader 3 | 99 | 93 | 0.55 | 80 | 50 | <1$^e$-3 |

**Supplementary Table 2.** Results of the quantitative evaluation of realistic value of the images by the three reviewers for the sub range $\sigma$ = 7-9 and $\sigma$ = 10-11.